\documentclass{llncs}
\usepackage{mathptmx}       
\usepackage{helvet}         
\usepackage{courier}        
\usepackage{type1cm}        
\usepackage{makeidx}         
\usepackage{graphicx}        
\usepackage{multicol}        
\usepackage[bottom]{footmisc}
\usepackage{subfigure}
\usepackage{amsfonts}
\usepackage[cmex10]{amsmath}

\begin{document}

\title{Vaccine skepticism detection by network embedding}
\titlerunning{Hamiltonian Mechanics}  
%
\author{Ferenc Béres \and Rita Csoma \and 
	Tamás Vilmos Michaletzky \and András A. Benczúr}
\authorrunning{Ferenc Béres et al.} 
%
\tocauthor{Ferenc Béres, Rita Csoma, Tamás Vilmos Michaletzky, András A. Benczúr}
\institute{Institute for Computer Science and Control, Hungary,\\
	\email{beres@sztaki.hu}
}
\maketitle    

\section{Introduction}

We compiled a data set to demonstrate the applicability of network embedding to vaccine skepticism, a controversial topic of long-past history that became more important than ever with the Covid-19 pandemic. Only a year
after the first international cases were registered, multiple vaccines were developed
and passed clinical testing.  Besides the challenges of development, testing and logistics, another factor in the fight against the pandemic are people who are hesitant to get vaccinated, or
even state that they will refuse any vaccine offered to them. Two groups of people
commonly referred to as
a) pro-vaxxer, those who support vaccinating people
b) vax-skeptic, those who question vaccine efficacy or the need for general vaccination against Covid-19.
It is very difficult to tell exactly how many people share each of these views. It
is even more challenging to understand all the reasoning why vax-skeptic opinions are
getting more popular.

In this work, our intention was to develop techniques that are able to efficiently differentiate between pro-vaxxer and vax-skeptic content. After multiple data preprocessing steps, we evaluated Twitter content and user interaction network classification by combining text classifiers with several node embedding and community detection models from an open-source Python library~\cite{karateclub}.
While several methods exist to embed by text content~\cite{tang2014learning} as well as by network structure~\cite{karateclub}, we are aware of only a few results that combine the two~\cite{yang2018enhanced,zhuo2019context,gong2020semi}.
Very recently, similar experiments~\cite{hui2021flipping} and a data set~\cite{muric2021vaccine} were published.

\textbf{Data.}
From January 7 to August 7, we collected data that anyone can view on Twitter by using the free Twitter API.
By using the keywords 
``vaccine'', ``vaccination'', ``vaccinated'', ``vaxxer'', ``vaxxers'', ``\#CovidVaccine'' and
``covid denier'', we collected 54,427 seed tweets, each with at least 50 replies (recursively). To eliminate drift towards topics of general politics such as US parties, we excluded the keywords ``Trump'',
``Biden'', ``republican'', ``democrat''. We considered seed tweets only in English. 
For each seed tweet, we collected the corresponding replies as well to build a reply network between Twitter users. In total, we collected almost $9$ million replies. 
In our experiments, we reduced the reply network to $579,159$ nodes and $4,156,502$ edges by dropping users with less than $3$ connections.

We randomly annotated $9.35$\% of the seed tweets in our data set with four different labels: pro-vaxxer ($2,626$ tweets), irrelevant ($1,438$ tweets), vax-skeptic ($681$ tweets) and anti-vaxxer ($344$ tweets). We found that it is even hard for humans to differentiate between vax-skeptic and anti-vaxxer content thus we merged anti-vaxxers into the vax-skeptic category. We trained a binary classifier over $2,626$ ($71.93$\%) pro-vaxxer and $1,025$ ($28.07$\%) vax-skeptic tweets by excluding irrelevant tweets.

\section{Results}

For classification, we used the following three modalities with logistic regression:
\begin{enumerate}
    \item \textbf{text:} $1,000$ dimensional TF-IDF vector of tweet text;
    \item \textbf{history:} Four basic statistics calculated from past tweet labels of the same user;
    \item \textbf{embedding:} $128$-dimensional user representation in the reply network.
\end{enumerate}

\begin{table}[t]
\centering
\begin{tabular}{ |l||r|r||r|r|| }
 \hline
  Feature components & AUC & gain (\%) & Accuracy & gain (\%) \\
 \hline
 text & 0.8385 & - & 0.7559 & -\\ 
 text+history & 0.8743 & 4.27 & 0.7769 & 2.78 \\  
 text+embedding & 0.9049 & 7.92 & 0.8427 & 11.48 \\ 
 text+embedding+history & 0.9130 & 8.88 & 0.8473 & 12.09 \\ 
 \hline
\end{tabular}
\caption{Model performance with different feature components. Performance gain is shown with respect to the simple baseline using only textual information. (AUC: Area Under the ROC Curve)}
\label{tab:results}
\end{table}

\begin{figure}[t]
\centering
\includegraphics[width=0.7\textwidth]{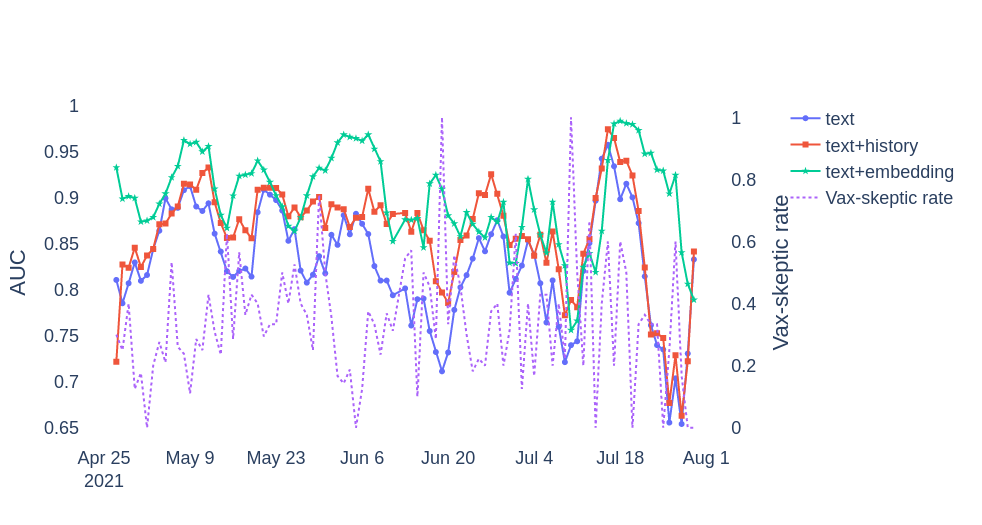}
\caption{Dynamic model performance based on a 7-day sliding window.}
\label{fig:temp_auc}
\end{figure}

\begin{figure}[t]
\centering
\includegraphics[width=0.7\textwidth]{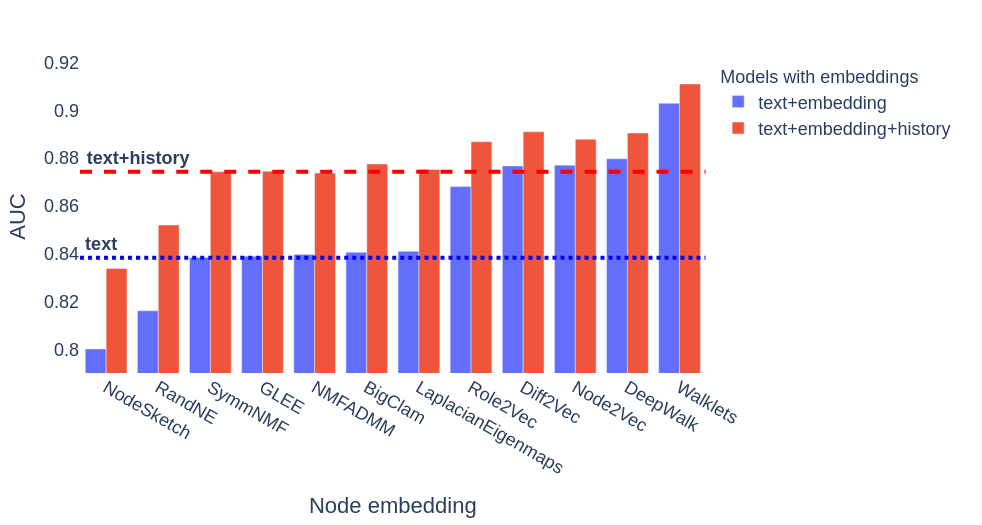}
\caption{Performance with respect to different node embedding and community detection models.}
\label{fig:node_emb_auc}
\end{figure}

\begin{figure}[t]
\centering
\includegraphics[width=0.3\textwidth]{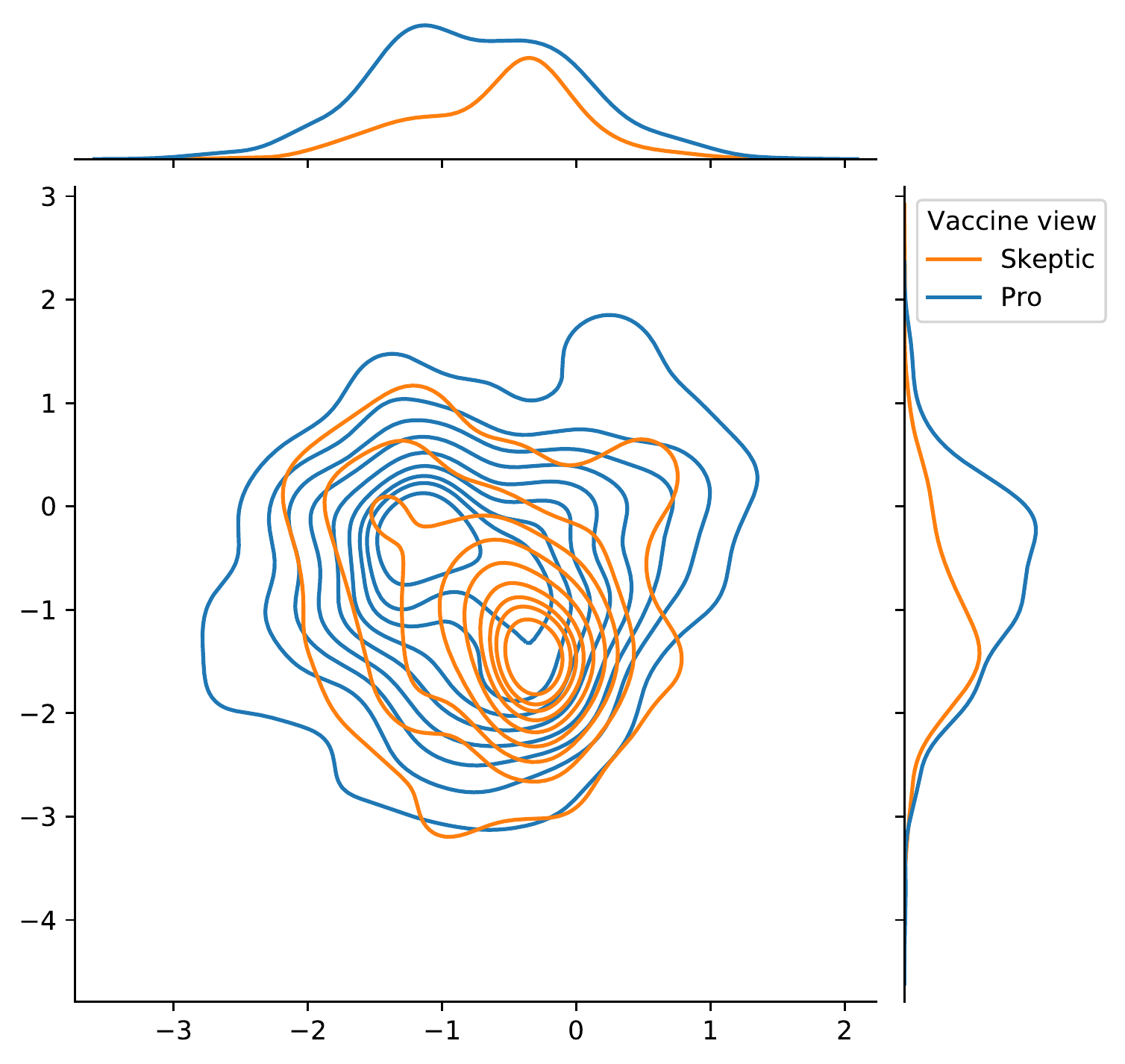}
\includegraphics[width=0.3\textwidth]{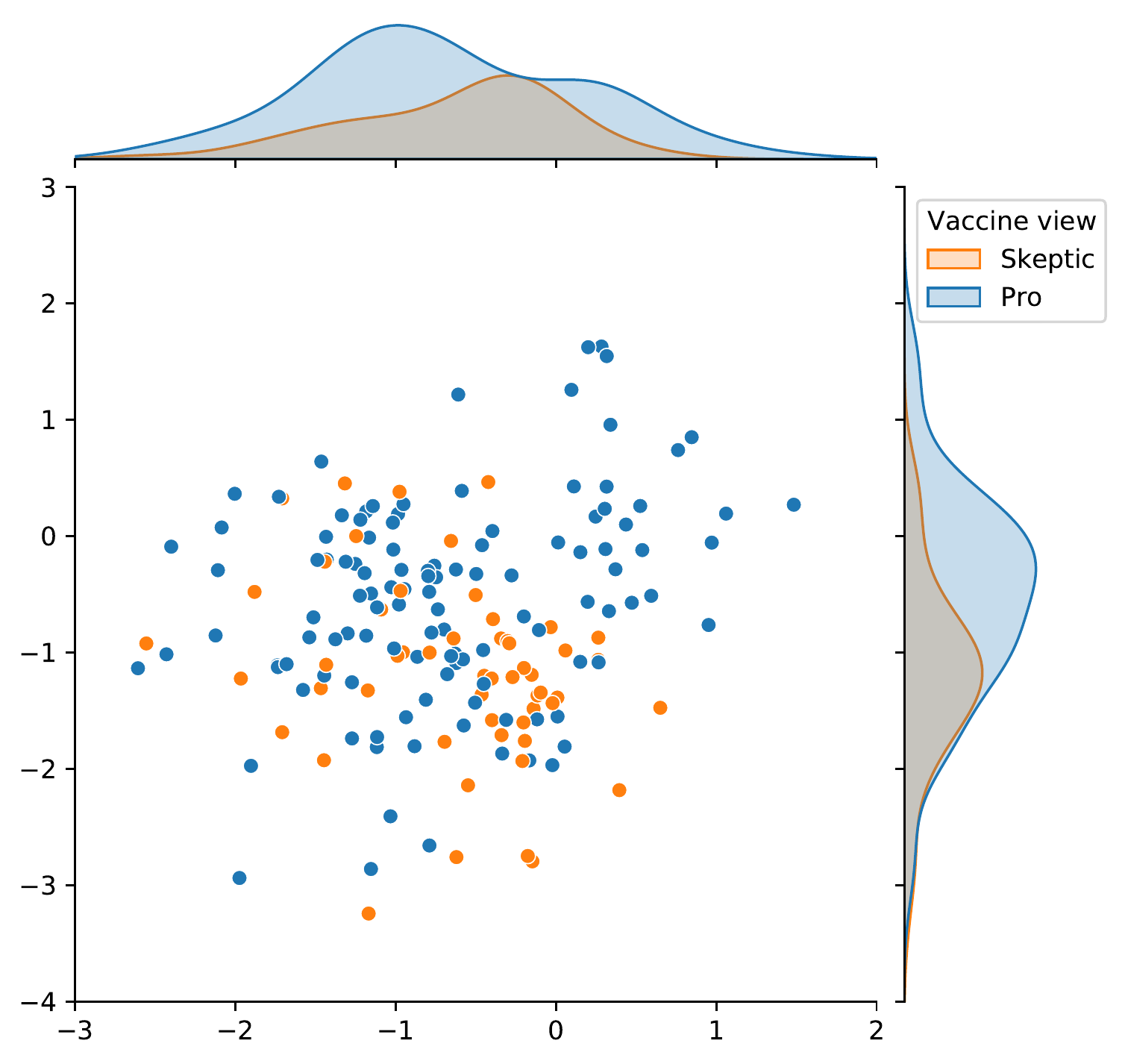}
\caption{Walklets clusters pro-vaxxer and vax-skeptic users well in the embedded space. \textbf{On the left} we show the kernel density estimation of the two groups for the whole test period, while  \textbf{on the right} only active users between 5-13 May are visualized.}
\label{fig:skeptic_clusters}
\end{figure}

We split the tweet data in time to 70\% training and 30\% testing.
Our results are summarized in Table~\ref{tab:results}. Not surprisingly, user statistics have a strong contribution as users usually stick to their past opinion.
User representations from the Twitter reply network improve performance, as seen in Figure~\ref{fig:temp_auc}. Indeed, tweets posted by users with no past label could be better inferred based on their social relations. Walklets~\cite{walklets}, the best performing node embedding model in Figure~\ref{fig:node_emb_auc}, even managed to find pro-vaxxer and vax-skeptic user clusters, see Figure~\ref{fig:skeptic_clusters}. For future work, we will replace the logistic regression classifier with a unified back-propagation neural network.

\textbf{Summary.}
In this work, we quantitatively showed that social interactions play a major role in  detecting vaccine skepticism. By deploying multiple node embedding models on a large Twitter reply network, we managed to discover pro-vaxxer and vax-skeptic communities.

For reproducibility and future research purposes, we share our data on GitHub\footnote{\url{https://github.com/ferencberes/covid-vaccine-network}}. In order to comply with the data publication policy of Twitter, we only share the user ID, original and reply tweet IDs along with the encoded content vectors.

\bibliographystyle{splncs}
\bibliography{abstract}

\end{document}